\documentclass[12pt,preprint]{aastex}

\begin{document}

\newcommand{\vsubmapr}{16.2}
\newcommand{\vXi}{1.51}
\newcommand{\vmaxSsig}{40}

\title{First Constraints on Source Counts at 350 Microns}

\author{Sophia A. Khan\altaffilmark{1,2,3}, Richard A. Shafer\altaffilmark{2}, Stephen Serjeant\altaffilmark{4}, S. P. Willner\altaffilmark{5}, Chris P. Pearson\altaffilmark{6, 7}, Dominic J. Benford\altaffilmark{2}, Johannes G. Staguhn\altaffilmark{2,8}, S. Harvey Moseley\altaffilmark{2}, Timothy J. Sumner\altaffilmark{3}, Matthew L. N. Ashby\altaffilmark{5}, Colin K. Borys\altaffilmark{9}, Pierre Chanial\altaffilmark{3}, David L. Clements\altaffilmark{3}, C. Darren Dowell\altaffilmark{10}, Eli Dwek\altaffilmark{2}, Giovanni G. Fazio\altaffilmark{5}, Attila Kov\'acs\altaffilmark{11}, Emeric Le Floc'h\altaffilmark{12,13}, Robert F. Silverberg\altaffilmark{2}}
\altaffiltext{1}{ALMA Fellow, Pontificia Universidad Cat\'{o}lica, Departamento de Astronomia y Astrofisica, 4860 Vicu\~{n}a Mackenna Casilla 306, Santiago 22, Chile}
\altaffiltext{2}{Observational Cosmology Laboratory (Code 665), NASA / Goddard Space Flight Center, Greenbelt, MD 20771}
\altaffiltext{3}{Imperial College of Science, Technology \& Medicine, Blackett Laboratory, Prince Consort Road, London SW7 2AZ, UK}
\altaffiltext{4}{Department of Physics and Astronomy, Open University, Walton Hall, Milton Keynes MK7 6AA, UK}
\altaffiltext{5}{Harvard-Smithsonian Center for Astrophysics, 60 Garden Street, Cambridge, MA 02138}
\altaffiltext{6}{Institute of Space and Astronautical Science, Yoshinodai 3-1-1, Sagamihara, Kanagawa 229 8510, Japan}
\altaffiltext{7}{ISO Data Centre, European Space Agency, Villafranca del Castillo, P.O. Box 50727, 28080 Madrid, Spain}
\altaffiltext{8}{Department of Astronomy, University of Maryland, College Park, MD 20742-2421}
\altaffiltext{9}{University of Toronto, Department of Astronomy \& Astrophysics, 60 George Street, Toronto, ON, Canada, M5S 3H4}
\altaffiltext{10}{Jet Propulsion Laboratory, California Institute of Technology, MC 169-327, 4800 Oak Grove Drive, Pasadena, CA 91109}
\altaffiltext{11}{Max-Planck-Institut f\"ur Radioastronomie, Auf dem H\"ugel 69, D-53121 Bonn, Germany}
\altaffiltext{12}{{\it Spitzer} Fellow, Institute for Astronomy, University of Hawaii, 2680 Woodlawn Drive, Honolulu, HI 96815, USA}
\altaffiltext{13}{Steward Observatory, University of Arizona, 933 N. Cherry Avenue,  Tucson, AZ 85721}

\begin{abstract}
We have imaged a $\sim$6\,arcminute$^2$ region in the Bo\"otes Deep Field using the 350\,$\mu$m-optimised second generation Submillimeter High Angular Resolution Camera (SHARC~II), achieving a peak 1$\sigma$ sensitivity of $\sim$5\,mJy.  We detect three sources above 3$\sigma$, and determine a spurious source detection rate of 1.09 in our maps.  In the absence of $5\sigma$ detections, we rely on deep 24\,$\mu$m and 20\,cm imaging to deduce which sources are most likely to be genuine, giving two real sources.   From this we derive an integral source count of 0.84$^{+1.39}_{-0.61}$\,sources~arcmin$^{-2}$ at $S>13$\,mJy, which is consistent with 350\,$\mu$m source count models that have an IR-luminous galaxy population evolving with redshift.  We use these constraints to consider the future for ground-based short-submillimetre surveys.  
\end{abstract}


\section{INTRODUCTION}

Towards the end of the last decade, a new population of submillimeter-selected galaxies (SMGs) were discovered through pioneering lensed and blank surveys using the 850\,$\mu$m-optimised Submillimetre Common User Bolometer Array (SCUBA; \citealt{Holland1999}) at the James Clerk Maxwell Telescope (\citealt{Smail1997}; \citealt{Hughes1998}; \citealt{Barger1998}; \citealt{Eales1999}). Extensive follow-up observations constrained the population to be mainly massive star forming galaxies (see, e.g., \citealt{fox02}; \citealt{Blain2002};  \citealt{Borys2003}) at high redshift \citep[z$\sim$2;][]{Chapman2005}, with the bulk of the luminosity emitted in the restframe far-IR --- although the detection of these sources had not been predicted by semi-analytical hiararchical models (for instance, contrast the order of magnitude spread in the models of \citealt{Guiderdoni1997} with the SCUBA-constrained single model of \citealt{Guiderdoni1998A}). SMGs can be considered the high redshift counterparts of the luminous and ultraluminous infrared galaxies (LIRGs and ULIRGs\footnote{Defined by $\rm 10^{11}\,L_\odot< L_{IR}~(8$--$\rm 1000\,\mu m)~<10^{12}\,L_\odot$ and $\rm L_{IR}~(8$--$\rm 1000\,\mu m)~>10^{12}\,L_\odot$ respectively}) found in the local universe (the majority selected with the Infrared Astronomical Satellite, {\it IRAS}; \citealt{Soifer1984}; \citealt{JosephWright1985}; \citealt{Soifer1987}).

The first surveys with SCUBA paved the way for many similar surveys using other submillimetre detectors (e.g., MAMBO, BOLOCAM, etc., \citealt{Bertoldi2000}; \citealt{Laurent2005}). These surveys were mostly limited to long submillimetre through millimetre wavelengths (500-1300\,$\mu$m), tracing emission on the long wavelength side of the peak at typical redshifts. The commissioning of the 350\,$\mu$m-optimised Second Generation Submillimeter High Angular Resolution Camera \citep[SHARC~II;][]{Dowell2003} at the Caltech Submillimeter Observatory (CSO), currently the largest ground-based submillimetre bolometer array \citep{Moseley2004}, provided a feasible opportunity to carry out a  blind survey in this waveband. 

Using SHARC~II, we targeted a $\sim 6$\,arcminute$^2$ region in the Bo\"otes Deep Field \citep{deVries2002} for a blank, deep survey, which was designed to select LIRGs and ULIRGs through their far-IR thermal dust emission (measured near the peak of the spectral energy distribution (SED)) at $1<z<3$, the epoch of peak cosmic star formation rate density \citep[see, e.g.,][]{Hopkins2006}. The survey, achieving a peak 1$\sigma$ sensitivity of $\sim$5\,mJy, produced a promising result as we reported the discovery of the first galaxy selected at 350\,$\mu$m \citep[SMM J143206.65+341613.4, also named Short Submillimetre Galaxy 1 (SSG~1);][]{Khan2005}.  
 
The discovery of SSG~1 raises a number of questions regarding the nature of galaxies detected in the short-submillimetre wavebands (200-500\,$\mu$m).  Given the demanding observational requirements \citep[good 350\,$\mu$m atmospheric transmission from Mauna Kea is $\sim$30\%\, as opposed to $\sim$80\%\ at 850\,$\mu$m;][]{Serabyn1998}, short-submillimetre surveys would be a poor use of ground-based telescope time if  they traced the same population as did long-submillimetre surveys. However, follow-up 1.2\,mm imaging appears to confirm the assertions of \citet{Khan2005}, that deep short-submillimetre observations can probe SMGs too faint for selection in longer submillimetre bands (faint SMGs), whose global properties might differ from the bright SMG population (e.g., lower redshift, warmer dust temperatures, lower luminosities; see \citealt{Khan2007}).  Given the paucity of 350\,$\mu$m-selected sources, the most efficient way to characterise the nature of the population is through deriving source counts and analysing the models that best fit the data. This complements the multiwavelength analysis on individual sources that was begun in other survey publications (\citealt{Khan2005}; \citealt{Khan2006}; \citealt{Khan2007}).

In this paper we present the first constraints on the source counts at 350\,$\mu$m. We outline our observation programme design, data reduction and analysis methodology. We discuss the criteria for selecting candidate 350\,$\mu$m sources and the determination of the number of spurious sources in the map. From this we derive the measured source counts from the survey. We discuss how the counts reflect the nature of our sources and conclude with the implications for future blank surveys in the short-submillimetre wavebands. 

\section{OBSERVATION PROGRAMME}
\label{sec:depth}
Submillimetre surveys have followed three approaches:  using gravitational lensing around clusters \citep[e.g.][]{Smail1997}, selecting fields surrounding known high redshift sources \citep[e.g., high redshift quasars;][]{Ivison2000}, and targeting a region of blank sky \citep[e.g.,][]{Hughes1998}.  For a given integration time, the number of detected sources will be higher in a lensing cluster survey as compared to a blank survey due to the brightness magnification.  However, this approach is highly dependent on the cluster mass distribution, which can produce significant systematic uncertainties on the luminosity function of the detected population and its evolution.  Even in the best possible case (a smooth cluster), imprecision in the cluster model could still dominate over the behaviour of the source counts. Submillimetre surveys centred on known high redshift sources run a risk of being redshift biased, since these are typically found at other wavelengths, and many are lensed. Additionally, correlation analyses show a higher probability of finding enhanced source counts over typical survey sizes in such areas \citep[see][and references therein]{Lagache2005}. 

To avoid the uncertainties associated with the biases listed above, we have chosen to pursue a blank survey, which can be implemented in ways ranging from deep, small area to shallow, large area surveys.  In order to maximise the number of detections in the survey, we could discriminate between the two approaches using the following argument:  the differential number versus flux relationship can be approximated locally as 
\begin{equation}
{N(S)} \approx k_d (S/S_0)^{-\gamma}~{\rm [sources~sr^{-1}~mJy^{-1}]}
\end{equation}
where $N(S)$ describes the overall surface density of galaxies as a function of flux density $S$. For a given limiting flux, $S_{min}$, the number of sources is
\begin{equation}
N(>S_{min}) = \int_{S_{min}}^{\infty} \frac{dN}{dS}dS\Rightarrow N(>S_{min})
\propto S_{min}^{1-\gamma}. 
\end{equation}
During a single pointed observation, the noise is expected to integrate down as $1/\sqrt{t}$. Hence the number of detected sources, $N$, is related to the integration time $t$ via
\begin{equation}
N_{\rm deep} \propto t^{(\gamma-1)/2}.
\end{equation} 
If the integration time was instead sub-divided into an equal number of shallower observations, this would yield
\begin{equation}
N_{\rm wide} \propto t. 
\end{equation}

Therefore a deep pointing yields more detections per exposure compared to a wider, shallow survey, as long as the flux density sensitivity remains at a level where $\gamma>$3.  For a non-evolving Euclidean universe $\gamma$=2.5, but current constraints on the submillimetre galaxy population show evolution ($\gamma>$2.5) for a broad range of brighter flux densities (e.g., \citealt{coppin06}). Constraining $\gamma$ through direct observation requires the detection of tens of sources at 350\,$\mu$m -- a huge demand on telescope time. 

Rather than parameterise the source counts from very small data sets, a more practical approach is to discriminate between existing source counts models, in particular those that successfully reproduce the IR-submillimetre counts.  Using the models in the literature at the time of the survey (\citealt{Franceschini1994}; \citealt{Guiderdoni1998B}; \citealt{Pearson2001}; \citealt{Takeuchi2001}) the target 1$\sigma$ sensitivity was based on where the models begin to show significant deviations in their source count predictions, with the majority of models having $\gamma>3$.  This threshold was 1$\sigma$=5\,mJy.

\subsection{Observations}

SHARC~II is a 350\,$\mu$m-optimized camera built around a $12\times 32$ element close-packed bolometer array. It achieves a point-source sensitivity of $\rm \sim 1\,Jy~s^{1/2}$ in good weather. The 384 pixels of the SHARC~II array image a region of around $1\farcm 0 \times 2\farcm 6$ on the sky. Its filled absorber array provides instantaneous imaging of the entire field of view, sampled at roughly 2.5 pixels per nominal beam area. 

The 350\,$\mu$m window is a difficult one for observers: the in-band atmospheric opacity $\tau$ is rarely $<0.8$, with signal-to-noise $S/N\propto e^{-\tau}/\sqrt{1 - e^{-\tau}}$, making efficient observations extremely weather dependent. For ground-based far-IR/submillimetre observations, the variation in atmospheric emission is the dominant noise source over all temporal frequencies.  Although rapid image differencing, commonly called chopping, is used to remove the atmosphere signal, this technique can give rise to a $\sqrt{2}$ increase in noise and a loss of observing time from a chopping duty cycle of $<1$. Furthermore, chopping does not adequately remove portions of the atmospheric signal that vary faster than the chop frequency,  something that our data reduction analysis has shown to exist \citep{Khan2006}. 

The design of SHARC~II eliminates the need to chop.  Atmospheric noise is spatially correlated, implying that the spatial variation in the atmosphere occurs in the line of sight of several pixels.  By scanning the detector array over the target region, the celestial signal -- spatially fixed and constant in time -- will be mapped by several detector pixels.  This scanning technique allows the determination of the individual pixel gains and offsets, and the removal of the atmospheric signal on all timescales; least squares fitting can also model other instrumental contributions, alongside the simultaneous derivation of the celestial sky map and associated uncertainty.  Although this modelling will induce some covariance between adjacent map pixels, this is small compared to the dominant contribution from photon noise. As part of the commissioning phase of SHARC~II, we tested a number of Lissajous scan patterns, typically using smaller amplitude sweeps of about 15\,arcseconds in the x-direction -- perpendicular to the 32 rows -- and 10-20\,arcseconds in the y-axis\footnote{The amplitude-period ratio should not be much larger than 1.4\,arcseconds per second.}.  This ensured that the entire area was well-covered, with substantial redundancy between detector pixels and map pixels.

The survey was awarded 12 half-nights of observing time, commencing in January 2003.  From that, just under seven hours of good quality data (from observations done in reasonable weather in January and March 2004) were obtained, centred on the Bo\"{o}tes Deep Field \citep{deVries2002} at position RA$=14^h32^m5\fs 75$, Dec$=34^\circ16'47\farcs 5$ (J2000), during the runs in January and March 2004. For these data the in-band zenith atmospheric opacity ($\tau_{350\,\mu\rm m}$) ranged from 1.0 to 1.4, corresponding to a zenith transmission of around 30 per cent. The beam profile was measured on known compact sources, and was verified to be within 3\%\ of the diffraction-limited beamwidth of $8.5''$. All observations were taken using the Dish Surface Optimisation System \citep{Leong2006}, which corrects for the primary mirror deformation as a function of zenith angle, to improve the telescope efficiency and the pointing.

\subsection{Data Reduction and Source Extraction}
\label{subsec:extract}

The data were reduced using the standard CSO reduction software, CRUSH
(\citealt{Kovacs2006}) version 1.40a8, using the advised reduction parameters for deep observations. This software implements a self-consistent least-squares algorithm to solve for the celestial emission, taking
into account instrumental and atmospheric contributions to the signal. Forty individual scans, each representing approximately ten minutes of integration time and all centred on the Bootes Deep Field position, were reduced simultaneously through CRUSH.  The output, the CRUSH-reduced skymap 
was calibrated with the flux density and point spread function based on observations of Callisto taken throughout the observing period at similar elevations (usually every hour).  The flux density of Callisto was derived from the CSO SHARC~II calibrator catalogue.  A thorough treatment of the reduction methodology, with detailed explanations of the reduction parameters, can be found in \citep{Khan2006}.

\begin{figure}[bthp]
\begin{center}
\plotone{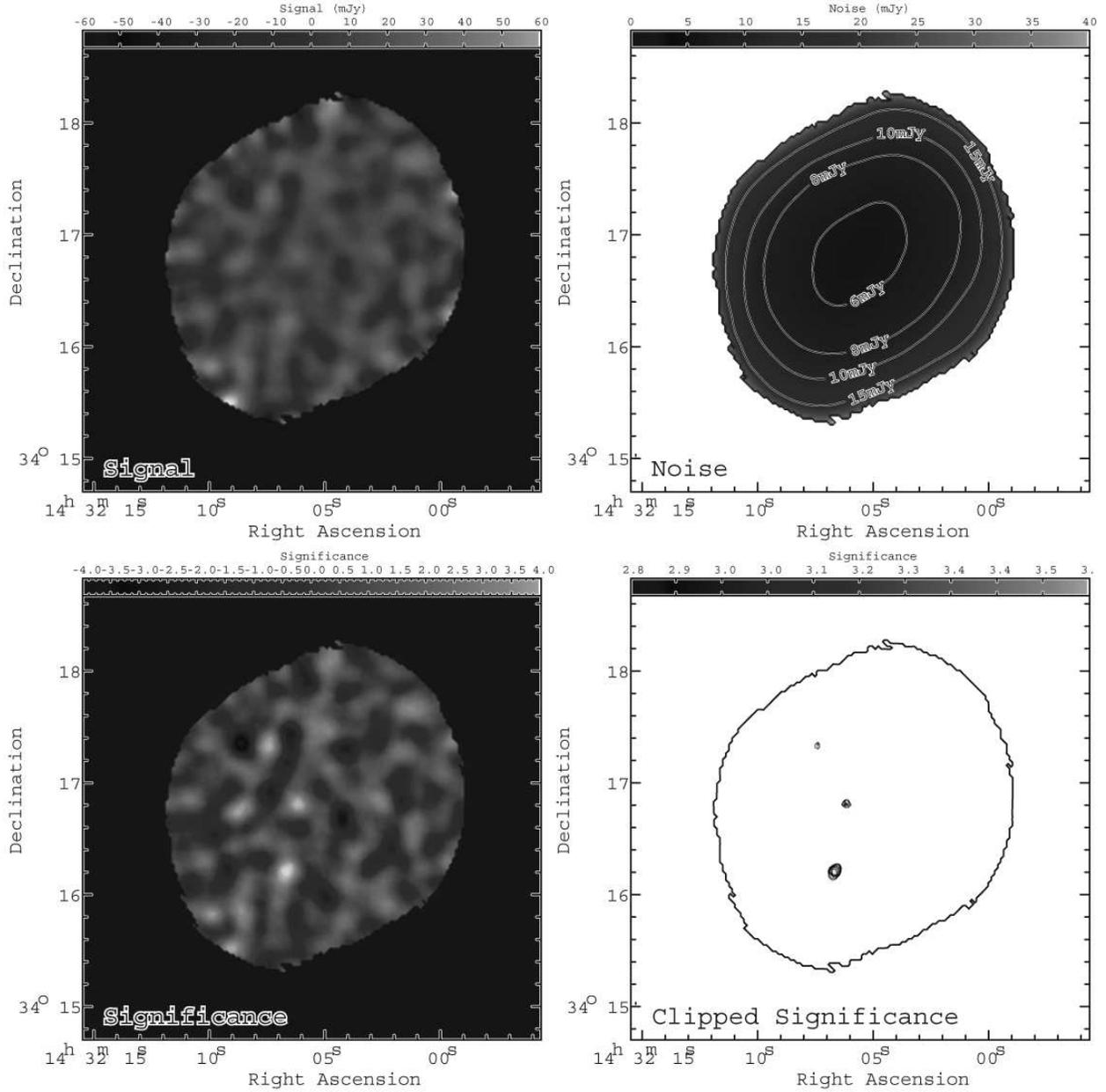}
\caption{Point source intensity {\it top left} and noise maps {\it top right} from the least square fit to the CRUSH reduced map [Jy per beam].  The bottom panels show the significance map {\it left}, and the map pixels with $\xi>2.8$ {\it right}.}
\label{figure:psfmaps}
\end{center}
\end{figure}

For each pixel in the CRUSH-reduced skymap, a least squares fit for a point source was determined. From the CRUSH celestial map, for each skymap pixel $j$, a sub-map comprising all pixels within \vsubmapr\ 
arcseconds (or 10 CRUSH skymap pixels), was extracted.  The size of the sub-map was chosen to provide a good determination of the source and background, but not so large as to require a more complicated background model, whereby four parameters were fit simultaneously: source intensity, mean background, and both a horizontal and vertical linear gradient. The Callisto point spread function (PSF) was then applied to this model in a weighted least squares fit --- this is roughly equivalent to smoothing the celestial map with the PSF (see Figure \ref{figure:psfmaps}).  For each pixel, this fit produces an intensity $S_{j}$ and an associated statistical uncertainty, $\sigma_{j}$, in units of flux density per beam.  These values allowed an estimate of the approximate signal-to-noise (S/N), which we refer to as the ``significance'' ($\xi_j$), using

\begin{equation}
    \xi_{j} = \frac{S_j}{\sigma_j}
    \label{eqn:eqxidef}
\end{equation}

This fitting reproduces the known 350\,$\mu$m flux densities of standard calibration sources to within the calibration uncertainties, but for faint sources, the map noise is the dominant uncertainty.

\subsection{Reweighting the map}
\label{subsec:xsnoise}

In a map with few detections, the expected distribution of $\xi$ will be Gaussian, with a variance of one, centered on zero.  We define
\begin{equation}
\Xi \equiv \sqrt{\frac{\sum_{j=0}^{N-1}\xi_{j}^{2}}{N}}
\end{equation}
as the rms variation in $\xi$.   

\begin{figure}[bthp]
\begin{center}
\plotone{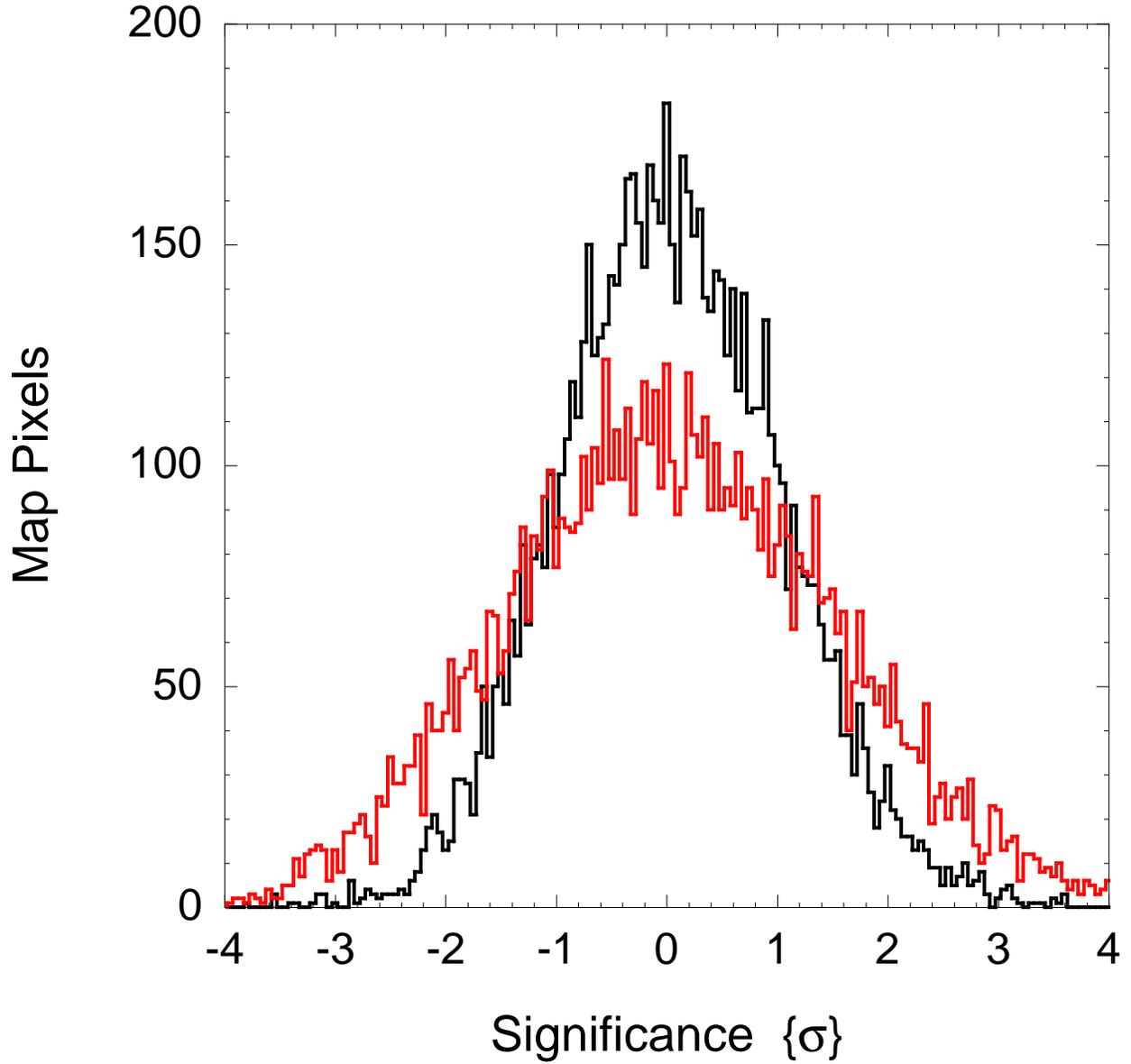}
\caption{Histogram of the uncorrected significance values, $\xi$ {\it red line}  (in units of $\sigma$), and corrected values, $\xi'$ {\it black line}, in the SHARC-Bootes map.  Scaling by $\Xi$ reduces the rms of the unscaled distribution to be $\approx$1.  }
\label{fig:gausshisto}
\end{center}
\end{figure}

\begin{figure}[bthp]
\begin{center}
\plotone{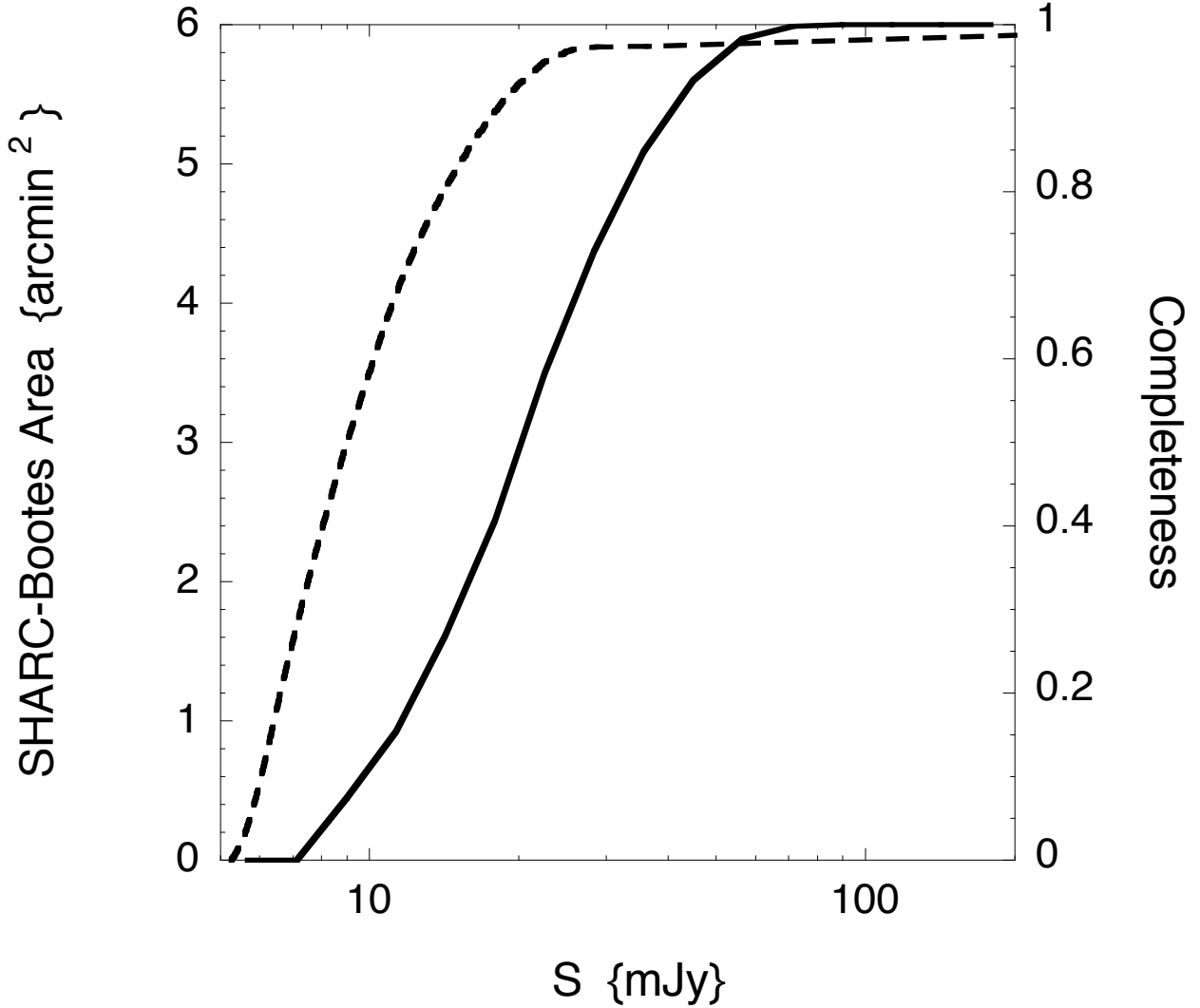}
\caption{Dashed-line: adjusted coverage map [arcmin vs 1$\sigma$ noise in Jy per beam] for the SHARC Bo\"{o}tes survey, after $\sigma$ is scaled by $\Xi$.  Thick-line: the completeness for the SHARC-Bootes survey, showing the fraction of simulated sources (with truth flux density, $S_{true}$ [mJy]) recovered at $\geq 3\sigma$.  We only consider the sources with 1$\sigma$ noise $\leq$10\,mJy ($A_{deep}=3.51$\,arcmin$^2$) thereby excluding the map edges.}
\label{fig:covmips}
\end{center}
\end{figure}

For the Bo\"otes data, $\Xi = \vXi$ -- implying further noise terms not accounted by in the CRUSH analysis.  While it is possible $\Xi>1$ could be due to real structure in the maps (such as confusion noise -- the statistical variation from unresolved sources), this is unlikely given the expected number of detections based on the survey sensitivity (using the models in Section \ref{sec:depth}).  Other models to derive an appropriate scaling factor were considered, from a simple constant offset to treating the excess noise as additional variance that is added in quadrature to the statistical uncertainty from the detector noise using maximum likelihood statistics \citep[see][]{Khan2006}, but an adequate solution was to simply scale the map by ${\Xi}$:  
\begin{equation}
    \sigma^{\prime}_{j} = \Xi \sigma_{j}
    \label{eq:sigmasp}
\end{equation}
giving a corrected significance of $\xi_{j}^{\prime} = \xi_{j} / \Xi$.  The magnitude of $\Xi$ appears stable with the integration time: real structure in the sky should be $\sqrt{t}$ more significant for longer integrations.  For source counts, the systematics associated with this excess noise are small compared to Poisson statistics.  From this point, $\xi$ and $\sigma$ refer to the adjusted values, $\xi'$ and $\sigma'$.  The adjusted significance distribution in the map is shown in Figure \ref{fig:gausshisto}, alongside the corresponding the survey coverage for the adjusted noise (Figure \ref{fig:covmips}).  It is this adjusted noise that is used for source extraction\footnote{Our analysis shows that this reweighting is still required in maps reduced with newer versions of CRUSH.}.

\subsection{Extracted Source Properties}

The corrected significance was used to select candidate detections, where $|\xi| \geq 3$.  There were three positive sources that met the detection criteria, including the previously reported SSG~1 (\citealt{Khan2005}; \citealt{Khan2006}), summarized in Table \ref{tab:bootessource} (note: $\sigma$ is scaled by $\Xi$), and two negative.  

The variation of $\chi^2$ with source position gives the position confidence contour, as given in Table \ref{tab:bootessource}, quoting 3$\sigma$ positional uncertainties (the best-fitting $\chi^2$ position will not necessarily match the peak S/N position, as illustrated by SSG~3 in Table \ref{tab:bootessource}).

\section{Constraints on the 350\,$\mu$m-selected population}

The relation between the measured density of sources and the corresponding flux densities (the source counts) constrains theoretical models of the source luminosity function and its evolution.  A thorough treatment of the measured counts would include a variety of statistical processes (e.g., confusion noise, errors in the map).  But the small number of detections in this survey means Poisson noise is dominant.  

Even in the absence of real sources ($\mu_S$), there will be still a statistical chance of detecting a source above the $\xi\geq  3\sigma$ threshold.  The mean number of these detections in the entire survey is called the accidental rate, $\mu_{A}$ (also referred to as the spurious source detection rate).  If the expected number of 350\,$\mu$m sources, both real and spurious, is small, then the two types of detections can be considered as independent detection processes, giving the total number of detections as $\mu = \mu_{A} + \mu_{S}$.

\subsection{Empirical estimate of the accidental rate}

A standard approach to determining the accidental rate is through using the pixel-pixel covariance to produce a model for the expected number of connected regions that lie above the detection threshold (3$\sigma$), assuming these covariances are well-characterised.  If the map noise obeyed Gaussian statistics, the probability of a pixel having $S/N >3\sigma$ per beam would be 0.00135. The approximate number of map beams is 310 (using the Callisto PSF).  Thus the expected number of accidental sources would be $\mu_{A} \approx 0.4$.  In the real CRUSH-reduced map, however, the difficulty in characterising the noise (Section~\ref{subsec:xsnoise}) shows that it is not Gaussian, which forces use of an alternative method for determining $\mu_{A}$.   One way is an empirical approach, similar to that used in \citet{Serjeant2003}, based on the fact that sky noise is not correlated with celestial position ($\alpha$, $\delta$) but real astronomical sources are.

For each raw data scan, a random rotation\footnote{Rotation angle is a parameter intended to represent the position angle of the SHARC~II array on the sky.  For present purposes, introducing a random value is nothing more than a simple method of offsetting the array astrometry from its true value.} angle was assigned, and the entire dataset with rotation angles was passed to CRUSH for reduction.  This has the effect of smearing the true astronomical sources while keeping the spatially correlated noise intact.  The source extraction method of Section~\ref{subsec:extract} was used to determine the number of candidate  sources in the rotated maps.  In total, 634 rotated maps were generated this way.  Although the corrupted-astrometry maps have  slightly different area-sensitivity coverage than the original map, the uncorrupted map is a random sample from this wider ensemble. The original map coverage is typical of the corrupted sample. The excess noise $\Xi$ of the original map is also within the range found for the corrupted maps (1.23-1.59).

The corrupted-astrometry maps produce the greatest density of spurious sources in the low-coverage, high-noise regions.  However, all the candidate 3$\sigma$ sources in Table~\ref{tab:bootessource} are in the central region, where $\sigma < 10$\,mJy). In this region, the spurious source detection rate is Poisson distributed with an expectation of $1.09\pm0.04$\footnote{The uncertainty is in the measurement of the accidental rate, not the range on the number of accidental sources.}

With three candidate point sources and an accidental rate of 1.09, the true detection rate is poorly determined.  However, observations at other wavelengths can assist in determining which sources are real.  Although this introduces a selection bias, it will be small compared to Poisson statistics.  Two of the candidate sources in Table~\ref{tab:bootessource} are 5$\sigma$ detections at 24\,$\mu$m and 20\,cm.  The probability of accidental detections at 24~\micron\ is 0.3 and 3\% for SSG~1 and 2 respectively.  At 20\,cm the accidental detection probability is 1\% for both sources.  Given these high-likelihood identifications it is unlikely that either of these two are spurious.   

SSG~3 is more problematic: the sensitivity of the 24\,$\mu$m data suffices to detect 850\,$\mu$m-selected galaxies (see, e.g., \citealt{Egami2004}).  The non-detection of this source at 24\,$\mu$m and 20\,cm suggests it is an atypical SMG, possibly at high redshift (see, e.g., \citealt{Ivison2002}; \citealt{Ivison2007}, although without the radio/24\,$\mu$m identification no photometric redshift estimate can be obtained).  But with the expectation of 1.09 spurious sources and the multiwavelength identifications of SSG~1 and SSG~2, we assume that SSG~3 is least likely to be genuine and so exclude it from further analysis.  

\subsection{Survey completeness}

To determine the survey completeness the two real sources, SSG~1 and SSG~2 (or SMM J143206.65+341613.4 and SMM J143206.11+341648.4), were removed from the CRUSH-reduced skymap and a source of random intensity was inserted into the no-source skymap, randomly placed over the entire skymap area, $A$. The simulated-source map was then fit as in Section \ref{subsec:extract}, and the fraction of simulated sources recovered at $\geq 3\sigma$ was determined through a Monte Carlo simulation (with the noise scaled by the same $\Xi$ as the original map).  The completeness against simulated source flux density is shown in Figure \ref{fig:covmips}, for the deepest part of the map ($\sigma<10$\,mJy).

\subsection{Source Counts}

The number of sources detected by a survey in area $A$ to depth $S>S_{min}$ will be
\begin{equation}
N_{det} = A \int_{S_{min}}^\infty{N(S)\times C(S)~dS}  
\end{equation}
where $C(S)$ is the completeness within the survey area.  Typical source count models (e.g., those given in  Section \ref{sec:depth}) are well represented by power laws in flux density, as given by Equation~1.  Setting $N_{det}=2$, substituting Equation 1 for $N(S)$, and normalising the differential counts at $S_0=20$\,mJy gives $k_d \approx 0.035$~sources~arcmin$^{-2}$~mJy$^{-1}$.  The normalisation at 20\,mJy gives the least dependence of $k_d$ on $\gamma$ for the present survey, for less than 10\% variation for $2.5\le\gamma\le4.0$.

The uncertainties on $k_d$ are set by Poisson statistics.  For an observed count of two objects, the true counts are between 0.53 and 5.32 with 90\% confidence \citep{Gehrels1986}.  The uncertainty on $k_d$ scales directly with these values.  Equation 2 allows direct comparison with integral count models.  We choose $S_{min} = 13$\,mJy, again minimizing the dependence on $\gamma$ for the actual survey, and find 0.84$^{+1.39}_{-0.61}$~sources~arcmin$^{-2}$ with $S>13$\,mJy (as shown in Figure \ref{fig:khansmail}), quoting the 90\%\ confidence uncertainty.  The variation is $<$5\% for $2.5\le\gamma\le4.0$.

In a map with few 3$\sigma$ detections, a careful consideration of the Eddington bias must be applied \citep[e.g.,][]{Eddington1913}.  Because there are usually more sources immediately below the flux limit than immediately above it, more sources are scattered above this limit, by positive noise fluctuations, than are scattered downwards to below it. Therefore, sources close to but above the flux limit have measured flux densities biased high, on average.  But if we assume a form for the source counts, the effect of Eddington bias is implicitly corrected.  However the deboosted individual flux densities are given in \citet{Khan2007}. 

\section{DISCUSSION}

\begin{figure}[bthp]
\begin{center}
\plotone{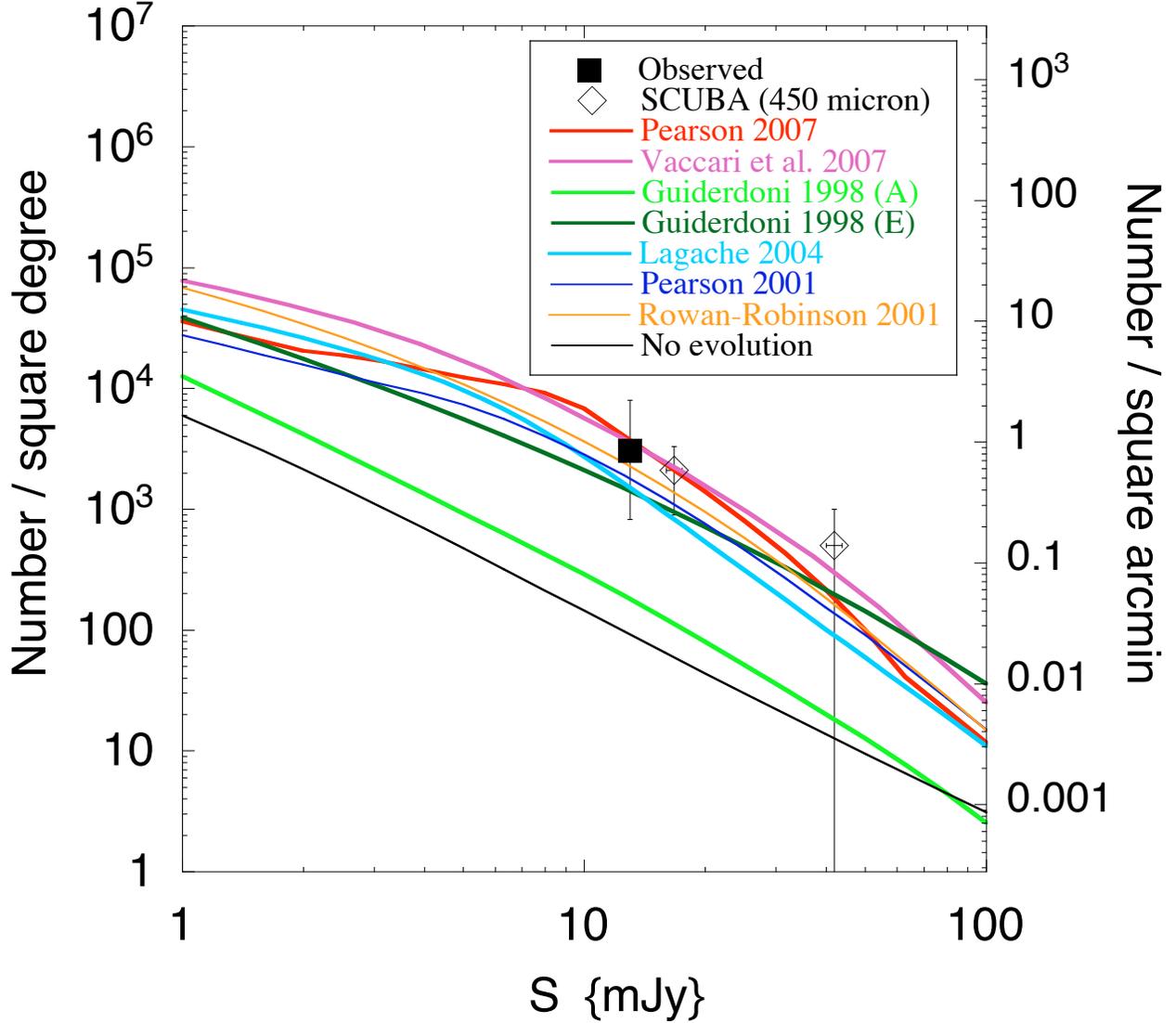}
\caption{The 350\,$\mu$m source count from this survey with corresponding 90\%\ confidence limits, and the shifted 450\,$\mu$m counts from \citet{smail02}.  The horizontal error bars on the 450\,$\mu$m counts give the range on the  Arp220 SED at $1<z<2$, with the actual flux density being the mean.  Thse counts are plotted with a selection of popular source count models from the literature, representing two methodologies of count modelling: backwards evolution (\citealt{Pearson2006}, \citealt{Vaccari2007}, \citealt{Lagache2004}, \citealt{Pearson2001}, \citealt{RR2001}) and semi-analytical (\citealt{Guiderdoni1998B}).  The \citet{Guiderdoni1998B} A model does not include a ULIRG population, whereas model E includes a strongly evolving ULIRG component within the extragalactic population.  Also shown is the No Evolution model of \citealt{Pearson2006}.}
\label{fig:khansmail}
\end{center}
\end{figure}

The derived integral counts are presented alongside a variety of source count models from the literature in Figure \ref{fig:khansmail}. The models represent two approaches to source count modelling -- backward evolution (\citealt{Pearson2006}, \citealt{Vaccari2007}, \citealt{Lagache2005}, \citealt{Pearson2001}, \citealt{RR2001}) and semi-analytic (\citealt{Guiderdoni1998B}) (see \citealt{HauserDwek01} for explanation and detailed descriptions of these methodologies).  The 350\,$\mu$m population, like other submillimetre-selected populations, is evolving with redshift, with numbers more than an order of magnitude higher than no-evolution predictions.  At 90\%\ confidence we are able to reject the No Evolution model, as well as the no-ULIRG model from \citet{Guiderdoni1998B}. But due to the small sample size the bulk of the 350\,$\mu$m models cannot yet be discriminated or rejected. 

The small area of this survey means the source counts will inevitably be affected by cosmic variance.  But the number of 5$\sigma$ 24\,$\mu$m detections within the SHARC-Bootes area compared to the counts of \citet{papovich04} suggest an underdensity in this field (see \citealt{Khan2007b}).  Also, the photometric redshifts of the two detected sources \citep[$z\sim1$ and $z\sim2$;][]{Khan2007} make it unlikely these objects are related to each other.

For comparison we plot the 450\,$\mu$m counts from \citet{smail02} in Figure \ref{fig:khansmail} assuming an Arp220 SED template to transform the 450\,$\mu$m counts to 350\,$\mu$m (the 450\,$\mu$m flux density of 10\,mJy being roughly equivalent to a 350\,$\mu$m flux density of $\sim$16\,mJy). Although this is a crude shift it appears consistent with the 350\,$\mu$m counts.  These counts are also consistent with the 350\,$\mu$m limits (at $\sim$25\,mJy) on 850\,$\mu$m-selected sources presented in \citet{Coppin2007}.  

Using the relation of \citet{Fixsen1998}, the 350\,$\mu$m contribution to the cosmic infrared background (see, e.g., \citealt{Lagache2005}) is 0.65\,MJy~sr$^{-1}$.   From the source counts we estimate resolving around 30\% of the 350\,$\mu$m background at 13\,mJy (with the entire 350\,$\mu$m background being resolved at a flux density of $\sim$0.5\,$\mu$Jy).  Although this is roughly the double the number resolved by the \citet{smail02} survey at $S_{350}$=16\,mJy (see also \citealt{Lagache2005}), the counts are extremely steep in this flux density domain and thus small increases in sensitivity result in large resolved fractions. 
 
\citet{Khan2007} discuss the spectral energy distributions of the two sources detected and show that the luminosities are $\sim 10^{12}$L$_\odot$ and dust temperatures are in the range 30--40\,K, placing them in the region of luminosity-dust temperature space between local IR-luminous galaxies and the colder, more luminous, and much more massive SCUBA sources \citet{blain04}.  This supports the argument of \citet{Khan2005} that the short-submillimetre might sample a warmer SMG population.  Indeed the upper limits at 1200\,$\mu$m (\citealt{Khan2007}) imply that the SHARC~II sources may lie below the detection limit of the SCUBA instrument at 850\,$\mu$m.  Given this survey resolves a larger fraction of the short-submillimetre background compared to the 850\,$\mu$m-bright sample of \citet{smail02}, it is possible faint SMGs outnumber SCUBA-bright sources (defining a faint SMG as $S_{850} \lesssim$5\,mJy).  

In order to better understand the nature of the short-submillimetre population it will be necessary to increase the number of sources, sampling a larger dynamic range in flux density.  This can be achieved through follow-up imaging of SMGs selected at long-submillimetre wavelengths (e.g., \citealt{kovacsetal06}; \citealt{Coppin2007}), or through deep surveys similar to this one.  But a far more efficient way will be through space-based and balloon-borne surveys. 

ESA's Herschel Space Observatory ({\it Herschel}, due for launch in $\sim$2008; \citealt{pilb02}; \citealt{harwit04}) will carry out both medium and deep surveys in the short-submillimetre wavelengths (250, 350 and 500\,$\mu$m) with the SPIRE instrument \citep[]{griffin04}. Similarly, the Balloon-borne Large Area Submillimeter Telescope \citep[BLAST;][]{devlin04} will conduct deep, large area surveys at submillimetre wavelengths including 350\,$\mu$m.  These surveys will select large numbers of sources, making it possible to assess the relative contribution of bright and faint SMGs to the short-submillimetre background, and determine, through multiwavelength analysis, whether the global properties of the short-submillimetre population are different from the SCUBA-bright SMG population. 

But the turnover in the 350\,$\mu$m differential counts is predicted to occur in the flux density range 5$<$ S$_{350}<$20\,mJy (e.g., \citealt{Lagache2004}; \citealt{Pearson2006}; \citealt{Vaccari2007}), which is below the 20 beams per source confusion limit ($\sim$21\,mJy) for the {\it Herschel} SPIRE wavebands (BLAST will also be confusion-limited at flux densities $\la$25\,mJy; \citealt{Pearson2006}).  This is a powerful diagnostic to discriminate both evolutionary models and the sources dominating the 350\,$\mu$m background --- sub-L* galaxies that will dominate the CIB and the volume-averaged star formation rate --- hence ultra-deep ground-based 350\,$\mu$m surveys could be the only plausible opportunity to detect this break for the foreseeable future, with the same argument applying to surveys in other short-submillimetre bands, e.g., 450\,$\mu$m with SCUBA~2 (\citealt{Holland2006}).

\section{CONCLUSION}

The SHARC-Bo\"otes survey is a $\sim$6\,arcminute$^2$ blank field survey that achieves a peak 1$\sigma$ 350\,$\mu$m sensitivity of $\sim$5\,mJy.  Having accounted for artificial sky structure in the map, we detect three candidate sources with S/N$\geq 3\sigma$.  From our three detections, we use a Monte Carlo simulation to deduce a spurious source detection rate, which is Poisson distributed with an expectation of 1.09 within the central region of the map.  Deep 24\,$\mu$m and 20\,cm imaging is used to confirm the detections and exclude spurious sources.  From this identification in other bands, and with a likelihood of one source being accidental, we believe there are two real 350\,$\mu$m-selected sources in our survey.  

Our source count indicates that the IR-luminous population at 350\,$\mu$m is evolving with redshift, with the no-evolution scenario rejected at 90\%\ confidence. 350\,$\mu$m surveys with BLAST, and after that, {\it Herschel}, may be unable to probe sources below our current survey detection threshold (due to the constraints of source confusion) where the differential counts are expected to turn over, therefore future ground-based observations should be designed to constrain this break through ultra-deep surveys.

\section{ACKNOWLEDGEMENTS}

We thank the anonymous referee for incisive comments that have improved this manuscript.  The Caltech Submillimeter Observatory is supported by NSF contract AST-0229008. We thank Tom Phillips and the CSO for observing time and assistance during our runs.  We are extremely grateful to Rick Arendt for his work on the GSFC SHARC~II data reduction effort.  We thank Dave Chuss for CSO observational support and Chris Carilli for assistance with the VLA observations and data reduction.    S.A.K. thanks David Hughes and Itziar Aretxaga for very helpful discussions.

S.A.K. thanks the following for funding support for this work: the Smithsonian Astrophysical Observatory, the Japan Society for the Promotion of Science, the Atacama Large Millimeter Array, the Comisi\'{o}n Nacional de Investigaci\'{o}n Cient\'{i}fica y Tecnol\'{o}gica de la Rep\'{u}blica de Chile and the Departamento de Astronomia y Astrofisica at Pontificia Universidad Cat\'{o}lica.  S.S. thanks PPARC for support under grant PP/D002400/1.  Support for E.L.F.'s work was provided by NASA through the Spitzer Space Telescope Fellowship Program.

\clearpage

\begin{table*}
\begin{center}
\begin{small}
\begin{tabular}{||c|c|c|c|c||} \hline \hline
Candidate Source & Flux density [mJy beam$^{-1}$] & Peak significance [$\sigma$] & Position [J2000]\\ \hline
SSG 1   & 23.2$\pm$6.5       & 3.6 & 14:32:06.65$\pm$0.26 +34:16:13.4$\pm$3.4 \\
SSG 2   & 17.1$\pm$5.4       & 3.2 & 14:32:06.11$\pm$0.28 +34:16:48.4$\pm$3.2 \\
SSG 3   & 19.9$\pm$7.1       & 3.0 & 14:32:07.46$\pm$0.39 +34:17:19.3$\pm$8.1\tablenotemark{a} \\ \hline \hline
\end{tabular}
\end{small}
\tablenotetext{a}{This confidence region is affected by the close proximity of two spurious sources with -3.6 and -2.2$\sigma$.}
\caption{Flux densities at the best-fitting position for sources in the SHARC-Bo\"{o}tes Field with significance $\geq $3.0$\sigma$ (quoting 3$\sigma$ positional uncertainties).}
\label{tab:bootessource}
\end{center}
\end{table*}

\clearpage

\clearpage

\clearpage

\clearpage

%


\begin{thebibliography}{}
\bibitem[Barger et al.(1998)]{Barger1998}Barger, A. J., et al., 1998, Nature, 394, 248
\bibitem[Bertoldi et al.(2000)]{Bertoldi2000} Bertoldi, F., et al.\ 2000, \aap, 360, 92
\bibitem[Blain et al.(2004)]{blain04}Blain, A. W., Chapman, S. C., Smail, I., Ivison, R., 2004, ApJ, 611, 52
\bibitem[Blain et al.(2002)]{Blain2002}Blain, A.W., Smail, I., Ivison, R.J., Kneib, J.-P., Frayer, D.T., PhR, 369, 111
\bibitem[Borys et al.(2003)]{Borys2003}Borys, C., Chapman, S., Halpern, M., Scott, D., 2003, MNRAS, 344, 385
\bibitem[Chapman et al.(2005)]{Chapman2005}Chapman, S. C., Blain, A. W., Smail, I., Ivison, R. J., 2005, ApJ, 
622, 772
\bibitem[Coppin et al.(2007)]{Coppin2007}Coppin, K., et al., 2007, submitted to MNRAS
\bibitem[Coppin et al.(2006)]{coppin06}Coppin, K., et al., 2006, MNRAS, 372, 162
\bibitem[Devlin et al.(2004)]{devlin04}Devlin M.J. et al., 2004, SPIE, 5498, 42
\bibitem[Dwek(2007)]{Dwek2006}Dwek, E., 2007, in preparation
\bibitem[Fox et al.(2002)]{fox02}Fox M. et al., 2002, MNRAS, 331, 839
\bibitem[de Vries et al.(2002)]{deVries2002}de Vries, W. H., Morganti, R., Rttgering, H. J. A., Vermeulen, R.,van Breugel, W., Rengelink, R., Jarvis, M. J., 2002, AJ, 123, 1784
\bibitem[Dowell et al.(2003)]{Dowell2003}Dowell C.D., et al., 2003, SPIE, 4855, 73
\bibitem[Dunlop(2001)]{Dunlop2001}Dunlop, J.~S.\ 2001, New Astronomy Review, 45, 609 
\bibitem[Eales et al.(1999)]{Eales1999}Eales, S., et al., 1999, ApJ, 515, 518
\bibitem[Eddington(1913)]{Eddington1913}Eddington, A., 1913, MNRAS 73, 359
\bibitem[Egami et al.(2004)]{Egami2004}Egami, E., et al., 2004, ApJS, 154, 130
\bibitem[Eisenhardt et al.(2004]{Eisen2004}Eisenhardt, P.\ R.\ et al.\ 2004, \apjs, 154, 48
\bibitem[Elston et al.(2006)]{Elston2006}Elston, R. J., et al. 2006, ApJ, 639, 816
\bibitem[Fixsen et al.(1998)]{Fixsen1998}Fixsen, D.J., et al., 1998, ApJ, 508, 123
\bibitem[Franceschini et al.(1994)]{Franceschini1994}Franceschini, A., Mazzei, P., de Zotti, G., \& Danese, L., 1994, ApJ, 427, 140
\bibitem[Gehrels(1986)]{Gehrels1986}Gehrels, N., 1986, ApJ, 303, 336
\bibitem[Griffin et al.(1999)]{griffin04} Griffin M.L., Swinyard B.M., Vigroux L., 2004, Optical, Infrared and Millimeter Space Telescopes. Edited by Mather, John C. Proceedings of the SPIE, 5487, 413
\bibitem[Guiderdoni et al.(1997)]{Guiderdoni1997}Guiderdoni, B., Hivon, E., \& Bouchet, F.~R.\ 1997, Extragalactic Astronomy in the Infrared, 521
\bibitem[Guiderdoni(1998a)]{Guiderdoni1998A}Guiderdoni, B.\ 1998a, ASP Conf.~Ser.~146: The Young Universe: Galaxy Formation and Evolution at Intermediate and High Redshift, 146, 283
\bibitem[Guiderdoni et al.(1998b)]{Guiderdoni1998B}Guiderdoni, B., Hivon, E., Bouchet, F. R., Maffei, B., 1998b, MNRAS, 295, 877
\bibitem[Harwit(2004)]{harwit04}Harwit M., 2004, Adv. Space Res., 34, 568
\bibitem[Holland et al.(2006)]{Holland2006}Holland, W. S., et al., 2006, SPIE, 6275, 45
\bibitem[Holland et al.(1999)]{Holland1999}Holland, W. S., et al., 1999, MNRAS, 303, 659
\bibitem[Hauser \& Dwek(2001)]{HauserDwek01}Hauser, M.G., \& Dwek, E., 2001, ARA\&A, 39, 249
\bibitem[Hopkins \& Beacom(2006)]{Hopkins2006}Hopkins, A.M., Beacom, J.F., 2006, ApJ, 651, 1 
\bibitem[Hughes et al.(1998)]{Hughes1998}Hughes D. et al., 1998, Nature, 457, 616
\bibitem[Ivison et al.(2007)]{Ivison2007}Ivison, R.~J., et al., 2007, astro-ph/0702544
\bibitem[Ivison et al.(2002)]{Ivison2002}Ivison, R.~J., et al., 2002, MNRAS, 337, 1
\bibitem[Ivison et al.(2000)]{Ivison2000}Ivison, R.~J., Dunlop,J.~S., Smail, I., Dey, A., Liu, M.~C., \& Graham, J.~R.\ 2000, \apj, 542, 27 
\bibitem[Jannuzi \& Dey(1999)]{Jannuzi1999}Jannuzi, B. T., Dey, A., 1999, in "Photometric Redshifts and the 
Detection of High Redshift Galaxies", ASP Conference Series, Vol. 191, Edited by R. Weymann, L. Storrie-Lombardi, M. Sawicki, and R. Brunner. ISBN: 158381-017-X, p. 111
\bibitem[Joseph \& Wright(1985)]{JosephWright1985}Joseph R.D., Wright G.S., 1985, MNRAS, 214, 87
\bibitem[Khan et al.(2005)]{Khan2005}Khan, S.A. et al. 2005, ApJ 631, L9
\bibitem[Khan(2006)]{Khan2006} Khan, S.A., 2006, PhD Thesis, University of London {\it available in electronic format on request}
\bibitem[Khan et al.(2007)]{Khan2007}Khan, S.A. et al. 2007a, in preparation
\bibitem[Khan et al.(2007)]{Khan2007b}Khan, S.A. et al. 2007b, in preparation
\bibitem[Kov\'acs et al.(2006)]{kovacsetal06}Kov\'acs, A., Chapman, S.C., Dowell,C.D., Blain, A.W.,Ivison, R.J., Smail, I., Phillips, T.G., 2006, MNRAS, 650, 592
\bibitem[Kov\'acs(2006)]{Kovacs2006}Kov\'acs, A., 2006, PhD Thesis, Caltech
\bibitem[Lagache et al.(2005)]{Lagache2005}Lagache, G., Puget, J.-L. \& Dole, H. 2005, ARAA, 43, 727
\bibitem[Lagache et al.(2004)]{Lagache2004}Lagache, G., et al., 2004, ApJS, 154, 112
\bibitem[Laurent et al.(2005)]{Laurent2005}Laurent, G.~T., et al.\ 2005, \apj, 623, 742
\bibitem[Leong et al.(2006)]{Leong2006}Leong, M., Peng, R., Houde, M., Yoshida, H., Chamberlin, R., \& Phillips, T.~G.\ 2006, \procspie, 6275,21
\bibitem[Mortier et al.(2005)]{Mortier2005}Mortier, A., et al., MNRAS 363, 563
\bibitem[Moseley et al.(2004)]{Moseley2004}Moseley S.H., Allen C.A., Benford D., Dowell C.D., Harper D.A.,Phillips T.G., Silverberg R.F., Staguhn J., 2004, NIMPA, 520, 417
\bibitem[Papovich et al.(2004)]{papovich04}Papovich C. et al.,  2004, ApJS, 154, 70
\bibitem[Pearson(2007)]{Pearson2006}Pearson, C.P., 2007, in preparation
\bibitem[Pearson(2001)]{Pearson2001}Pearson, C.P., 2001, MNRAS, 325, 1511
\bibitem[Pilbratt(2002)]{pilb02}Pilbratt G. L., 2002, in Mather J.C., ed., Proc. SPIE Vol. 4850, IR Space Telescopes and Instruments. SPIE, Bellingham, WA, p. 586
\bibitem[Rowan-Robinson(2001)]{RR2001}Rowan-Robinson, M., 2001, ApJ, 549, 745
\bibitem[Schuster et al.(2006)]{2006SPIE.6270E..65S} Schuster, M.~T., Marengo, M., \& Patten, B.~M.\ 2006, \procspie, 6270
\bibitem[Serabyn et al.(1998)]{Serabyn1998}Serabyn, E., Weisstein, E.~W., Lis, D.~C., \& Pardo, J.~R.\ 1998, \ao, 37, 2185 
\bibitem[Serjeant et al.(2003)]{Serjeant2003}Serjeant, S., et al., 2003, MNRAS, 344, 887
\bibitem[Smail, Ivison \& Blain(1997)]{Smail1997}Smail I., Ivison R.J.,Blain A.W., 1997, ApJ, 490, L5
\bibitem[Smail et al.(2002)]{smail02}Smail I., Ivison R.J., Blain A.W., Kneib J.-P., 2002, MNRAS, 331, 495
\bibitem[Soifer et al.(1984)]{Soifer1984}Soifer B.T., et al., 1984, ApJ, 278L, 71
\bibitem[Soifer et al.(1987)]{Soifer1987}Soifer B.T., Neugebauer G., Houck J.R., 1987, ARA\&A, 25, 187
\bibitem[Takeuchi et al.(2001)]{Takeuchi2001}Takeuchi, T. et al. 2001, PASJ, 53, 37
\bibitem[Vaccari et al.(2007)]{Vaccari2007}Vaccari, M., et al., 2007, in preparation

\end{thebibliography}
\end{document}